# A Physically Based Analytical Model to Predict Quantized Eigen Energies and Wave Functions Incorporating Penetration Effect

Nadim Chowdhury, Imtiaz Ahmed, Zubair Al Azim, Md. Hasibul Alam, Iftikhar Ahmad Niaz and Quazi D.M. Khosru

Department of Electrical and Electronic Engineering

Bangladesh University of Engineering and Technology, Dhaka-1000, Bangladesh

We propose a physically based analytical compact model to calculate Eigen energies and Wave functions which incorporates penetration effect. The model is applicable for a quantum well structure that frequently appears in modern nano-scale devices. This model is equally applicable for both silicon and III-V devices. Unlike other models already available in the literature, our model can accurately predict all the eigen energies without the inclusion of any fitting parameters. The validity of our model has been checked with numerical simulations and the results show significantly better agreement compared to the available methods.

## Introduction

With continuous aggressive scaling of silicon devices in order to meeting up the challenge of Moore's law, MOSFET technology has reached a point where it has seen the introduction of channel strain, high-k dielectric, fully depleted Silicon on insulator (SOI) and Ultra thin box and body (UTBB) devices (1-3). The features of these devices include: improvement of the channel mobility through strain, strong control of gate on channel potential through high-k dielectric, and reduction of short channel effects (SCE) through full-depletion of the body in SOI (4-5). On the other hand, III-V materials such as GaAs, GaSb, InAs etc. offer low effective mass for conduction compared to silicon and thus it can be an alternative to silicon. Excellent features such as high mobility, low Effective mass and high saturation velocity of these III-V materials facilitate devices like HEMT, MOS-HEMT, QW-FET etc. to be very high speed devices. Moreover, different device structures has also been proposed for silicon for better electrostatics and to reduce SCE such as Double gate, Triple gate, G-4 FET, Gate all around, ETB (Extremely Thin Body) devices and many more.

In order to accurately model these various nano-scale devices, an accurate method must be available for the determination of eigen energies and wave functions of the Quantum-well structures present within these devices. The accurate prediction of eigen energies and wave functions, incorporating penetration (6) is of tremendous importance for the calculation of charge distribution in the channel & compact modeling of C-V and I-V characteristics (7-9).

Here, we present a physically based rigorous mathematical analysis of the quantum well that is frequently observed in the state-of-art silicon and III-V material based devices. Through this analysis, we propose a compact analytical model for the estimation of eigen energies and wave functions. The figure below (Fig. 1) shows the Quantum well that has been put under extensive investigation.

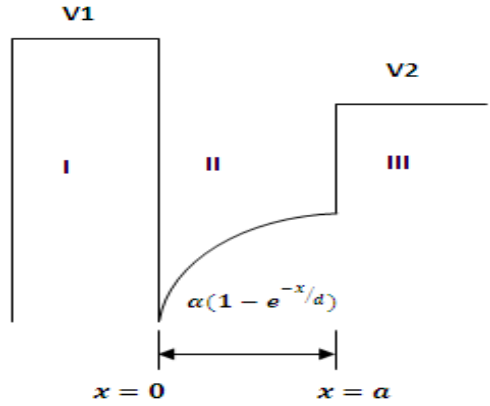

Figure 1: Quantum well under investigation

## Model Development

V1, $\alpha(1 - e^{-x/d})$ and V2 are the assumed potential for the three regions in the structure of Fig 1. Solving Schrodinger's equations, wave functions for the three regions can be represented as following.

$$\psi_1(x) = Ae^{\lambda x}u[-x] \tag{1}$$

$$\psi_2(x) = \left[B\,J_R\left(s\,e^{-x/2d}\right) + C\,J_{-R}\left(s\,e^{-x/2d}\right)\right](u[x] - u[x-a]) \tag{2}$$

$$\psi_3(x) = De^{\gamma x}u[x-a] \tag{3}$$

Where, $J_R(x)$ is the Bessel Function, $u[x]$ is the heaviside function and parameters $\lambda, \gamma, R, s$ are defined as following:

$$\lambda = \sqrt{V_1 - E}/\mu \tag{4}$$

$$\gamma = \sqrt{V_2 - E}/\mu \tag{5}$$

$$R = \frac{2d}{\mu}\sqrt{\alpha - E} \tag{6}$$

$$s = \frac{2d}{\mu}\sqrt{\alpha} \tag{7}$$

Applying the appropriate Boundary conditions we get:

$$\lambda = \frac{\gamma P + N(R,a)N(-R,0) - N(-R,a)N(R,0)}{\gamma Q + N(R,a)\,J_{-R}(s) - N(-R,a)\,J_R(s)} \tag{8}$$

With,

$$P = [N(R,0) J_{-R}(s\, e^{-a/2d}) - N(-R,0) J_R(s\, e^{-a/2d})] \quad [9]$$

$$Q = [J_R(s) J_{-R}(s\, e^{-a/2d}) - J_{-R}(s) J_R(s\, e^{-a/2d})] \quad [10]$$

$$N(R,a) = \frac{d}{dx}[J_R(s\, e^{-x/2d})]_{x \to a} = \frac{s}{4d} e^{-a/2d}[J_{R+1}(s\, e^{-x/2d}) - J_{R-1}(s\, e^{-x/2d})] \quad [11]$$

Using the definitions we derive the following relations:

$$\gamma = \sqrt{\frac{V_2 - \alpha}{\mu^2} + \frac{R^2}{4d^2}} \quad [12]$$

$$\lambda = \sqrt{\frac{V_1 - \alpha}{\mu^2} + \frac{R^2}{4d^2}} \quad [13]$$

Equating eqn. (8) and eqn. (13) and solving graphically we obtain the intermediate parameter R. The graphical solution of this equation yield multiple values for R, each value corresponds to one eigen state. The highest value of R yields the lowest eigen state. Eigen states are found with the following relation.

$$E = \alpha - \frac{\mu^2 R^2}{4d^2} \quad [14]$$

From the four boundary conditions we have derived an identity and from that we have expressed A,B and D as a function of C. The following equations shows the explicit compact form of the wave function of the quantum well under investigation.

$$\psi(x) = Me^{\lambda x}u[-x] + Di * e^{\gamma x}u[x-a] + [\kappa * J_R(s\, e^{-x/2d}) + J_{-R}(s\, e^{-x/2d})](u[x] - u[x-a]) \quad [15]$$

Where

$$M = \kappa * J_R(s) + J_{-R}(s) \quad [16]$$

$$Di = e^{-\gamma x} * (\kappa * J_R(ss) + J_{-R}(ss)) \quad [17]$$

$$C_{04}\kappa^4 + C_{03}\kappa^3 + C_{02}\kappa^2 + C_{01}\kappa^1 + C_{00} = 0 \quad [18]$$

$$C_{00} = \Lambda^2 T^2 - \Gamma^2 Y^2 + \Lambda^2 Y^2 \Theta \quad [19]$$

$$C_{01} = 2\begin{pmatrix} \Lambda^2 ZT + \Delta\Lambda T^2 - \Gamma^2 \Phi Y - X\Gamma Y^2 \\ +\Lambda^2 \Phi Y\Theta + \Delta\Lambda Y^2 \Theta \end{pmatrix} \quad [20]$$

$$C_{02} = \Lambda^2 Z^2 + 4\Delta\Lambda ZT + \Delta^2 T^2 - \Gamma^2 \Phi^2 - 4X\Gamma\Phi Y - \Gamma^2 Y^2 + \Lambda^2 \Phi^2 + 4\Delta\Lambda \Phi Y\Theta + \Delta^2 Y^2 \Theta \quad [21]$$

$$C_{03} = 2\begin{pmatrix} \Delta\Lambda Z^2 + ZT\Delta^2 + X\Gamma\Phi^2 - X^2\Phi Y \\ +\Delta\Lambda\Phi^2\Theta + \Delta^2 \Phi Y\Theta \end{pmatrix} \quad [22]$$

$$C_{04} = \Delta^2 Z^2 - \Gamma^2 \Phi^2 + \Delta^2 \Phi^2 \Theta \quad [23]$$

$$X = N(R,0), \Gamma = N(-R,0) \quad [24]$$

$$\Delta = J_R(s), \Lambda = J_{-R}(s) \quad [25]$$

$$Z = N(R,a), T = N(-R,a) \quad [26]$$

$$\Phi = J_R(ss), Y = J_{-R}(ss) \quad [27]$$

$$\Theta = \frac{V1-V2}{\mu^2} \quad [28]$$

## Results and Discussions

To validate the proposed model, a self-consistent simulator (10) for DG-FET is designed which addresses the coupled Poisson's and Schrodinger equations. DG-FET has the quantum well under consideration. From the obtained conduction-band profile, the model parameters α and d are obtained through simple fitting. Here the parameters α and d depends upon channel electric field and doping concentration, for the same electric field with different doping concentration the value of these parameters can be different. Using obtained values of α and d, Eigen energies are calculated using proposed model at various top gate bias and channel thickness. These results are compared with the self-consistent simulation results which show excellent agreement (Table I and II).

TABLE I. Eigen energies with $\alpha = 0.165\ eV,\ d = 5.328\ nm$

| Eigen Energy | Self-Consistent Results (eV) | Proposed Model (eV) | % Error |
|---|---|---|---|
| 1st | 0.0671 | 0.0666 | 0.7% |
| 2nd | 0.1033 | 0.1047 | 1.3% |
| 3rd | 0.1269 | 0.1284 | 1.2% |

TABLE II. Eigen energies with $\alpha = 0.2051\ eV,\ d = 3.592\ nm$

| Eigen Energy | Self-Consistent Results (eV) | Proposed Model (eV) | % Error |
|---|---|---|---|
| 1st | 0.0998 | 0.0991 | 0.7% |
| 2$^{nd}$ | 0.1477 | 0.1485 | 0.5% |
| 3$^{rd}$ | 0.1756 | 0.1775 | 1.1% |

TABLE III. Eigen energies comparison with Airy function using $\alpha = 0.236\ eV, \quad d = 5.031\ nm$

| Eigen Energy | Self-Consistent Results (eV) | Airy function (eV) | Proposed Model (eV) |
|---|---|---|---|
| 1$^{st}$ | 0.09127 | 0.1048 | 0.0887 |
| 2$^{nd}$ | 0.1426 | 0.1844 | 0.1409 |
| 3$^{rd}$ | 0.1739 | 0.2493 | 0.1747 |
| 4$^{th}$ | 0.1966 | 0.3065 | 0.1983 |

The variation of different Eigen energies with the top gate bias is observed with a fixed channel thickness. Comparison with the self-consistent simulation results yields tremendous matching (Fig 2). Using the proposed model, the obtained wave functions $|\psi_1|^2$, $|\psi_2|^2$ and $|\psi_3|^2$ are plotted. The results are in excellent agreement with the self consistently obtained wave functions (Fig 3).

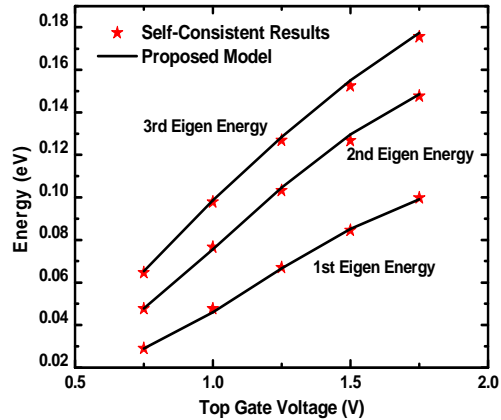

Figure 2. Comparison of eigen energies

With the approximation of $a = d$ and $V2 = \alpha$, the proposed model becomes applicable for surface channel MOSFETs (both Si and other materials). Table III shows the comparison among the values of eigen energies obtained from numerical simulation, proposed model and Airy function approximation (6-7) for a surface channel MOSFET. In all the cases, proposed model yields a closer value to the numerical one than the Airy function approximation. Airy function predicts the first eigen state with reasonable accuracy. However, for higher states this approximation produces very faulty results. In a comparative view our proposed model can accurately predict the higher states as well as the ground state with tremendous accuracy (Fig. 4).

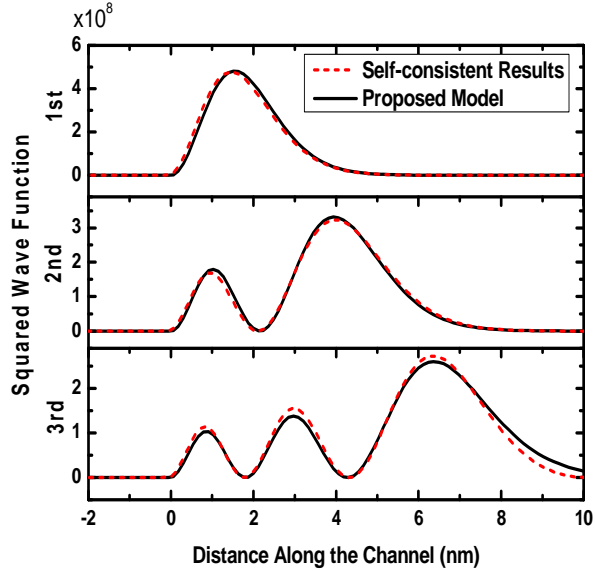

Figure 3. Comparison of wave functions.

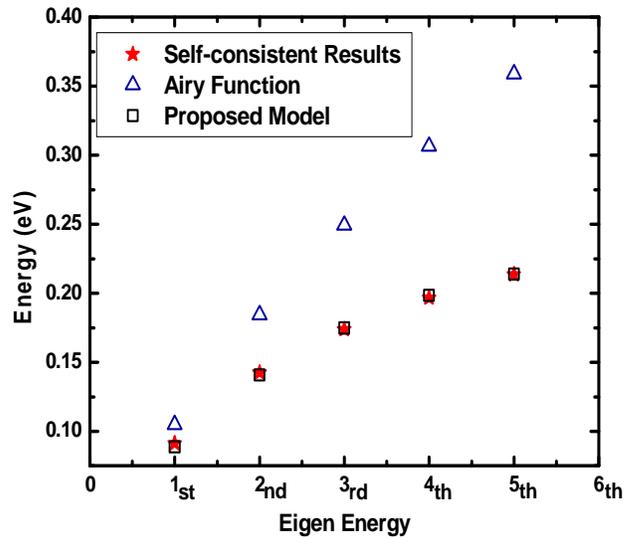

Figure. 4. Comparison of the proposed model with Airy function approximation

## Conclusion

An accurate physics based compact model to calculate sub-band energies and wave functions has been presented here. The accuracy of the proposed model vastly depends on the extraction parameters α and d, whereas the value of these two parameters depends on channel doping and electric field.

## Acknowledgement

Authors would like to thank Head, Department of EEE, Bangladesh University of Engineering and Technology for technical and financial support of this work.